\documentclass[twocolumn,twocolappendix]{aastex631}

\usepackage{tikz}
\usepackage{CJK}
\usepackage{multirow}

\usepackage{amsmath,amssymb,amstext}
\usepackage{graphicx}
\usepackage{enumitem}
\usepackage{natbib}

\DeclareMathAlphabet\mathbfcal{OMS}{cmsy}{b}{n}

\newcommand{\stag}{\ensuremath{\star}}
\newcommand{\J}{\ensuremath{\mathbf{J}}}
\newcommand{\ang}{\boldsymbol{\theta}}
\newcommand{\params}{\ensuremath{\alpha}}
\newcommand{\kpckms}{\ensuremath{\mathrm{kpc\,km\,s^{-1}}}}

\newcommand{\p}{\partial}

\newcommand{\bp}{\mathbf{p}}
\newcommand{\bq}{\mathbf{q}}

\newcommand{\prob}{\ensuremath{p}}

\newcommand{\tm}{$^{\mathsf{TM}}$\ }

\definecolor{mygreen}{HTML}{006d2c}
\definecolor{myblue}{HTML}{0868ac}
\definecolor{mybleuclair}{HTML}{f7fbff}
\definecolor{mypurple}{HTML}{6a51a3}
\definecolor{mybrown}{HTML}{67001f}
\definecolor{mypink}{HTML}{980043}
\definecolor{mypinkclair}{HTML}{f7f4f9}
\definecolor{myorange}{HTML}{fe9929}
\definecolor{myorangeclair}{HTML}{fff7ec}
\definecolor{mypurpleclair}{HTML}{fcfbfd}
\definecolor{mybluea}{HTML}{084081}

\shorttitle{Iron Snails and others}

\begin{document}
\author{Neige Frankel}
\affiliation{Canadian Institute for Theoretical Astrophysics, University of Toronto, 60 St. George Street, Toronto, ON M5S 3H8, Canada}
\affiliation{David A. Dunlap Department of Astronomy and Astrophysics, University of Toronto, 50 St. George Street, Toronto, ON M5S 3H4, Canada}

\author{David W. Hogg}
\affiliation{Max Planck Institute for Astronomy, K\"onigstuhl 17, D-69117 Heidelberg, Germany}
\affiliation{Center for Cosmology and Particle Physics, Department of Physics, New York University, 726~Broadway, New~York,~NY 10003, USA}
\affiliation{Flatiron Institute, 162 Fifth Avenue, New~York,~NY 10010, USA}

\author{Scott Tremaine}
\affiliation{Canadian Institute for Theoretical Astrophysics, University of Toronto, 60 St. George Street, Toronto, ON M5S 3H8, Canada}
\affiliation{School of Natural Sciences, Institute for Advanced Study, Princeton, NJ 08540, USA}

\author{Adrian Price-Whelan}
\affiliation{Center for Cosmology and Particle Physics, Department of Physics, New York University, 726~Broadway, New~York,~NY 10003, USA}
\affiliation{Flatiron Institute, 162 Fifth Avenue, New~York,~NY 10010, USA}

\author{Jeff Shen}
\affiliation{Department of Astrophysical Sciences, Princeton University, 4 Ivy Lane, Princeton, NJ 08544, USA}

\title{Iron Snails:\\ non-equilibrium dynamics and spiral abundance patterns}
%
%
%
%
%
%

\begin{abstract}\noindent
Galaxies are not in a dynamical steady state. They continually undergo perturbations, e.g., from infalling dwarf galaxies and dark-matter substructure. 
After a dynamical perturbation, stars phase mix towards a new steady state; in so doing they generally form spiral structures, such as spiral density waves in galaxy disks and the Gaia Snail\tm observed in the vertical phase-space density in the solar neighborhood.
Structures in phase-space density can be hard to measure accurately, because spatially varying selection effects imprint their own patterns on the density. However, stellar labels such as metallicity, or other element abundances, or stellar masses and ages, can be measured even in the face of complex or unknown spatial selection functions. We show that if the equilibrium galaxy has phase-space gradients in these labels, any perturbation that could raise a spiral wave in the phase-space density will raise a spiral wave in the distribution of labels as well.
We work out the relationship between the spiral patterns in the density and in the labels. As an example, we analyze the Gaia Snail\tm and show that its amplitude and dynamical age as derived from elemental abundances (mainly [Mg/Fe]) follow similar patterns to those derived from the phase-space density. Our best model dates the Snail's perturbation to about 400 Myr ago although we find significant variations with angular momentum in the best-fit age.
Conceptually, the ideas presented here are related to Orbital Torus Imaging, chemical tagging, and other methods that use stellar labels to trace dynamics. 
\end{abstract}


\section{Introduction}

\begin{figure*}
    \centering
    \includegraphics[width=\textwidth]{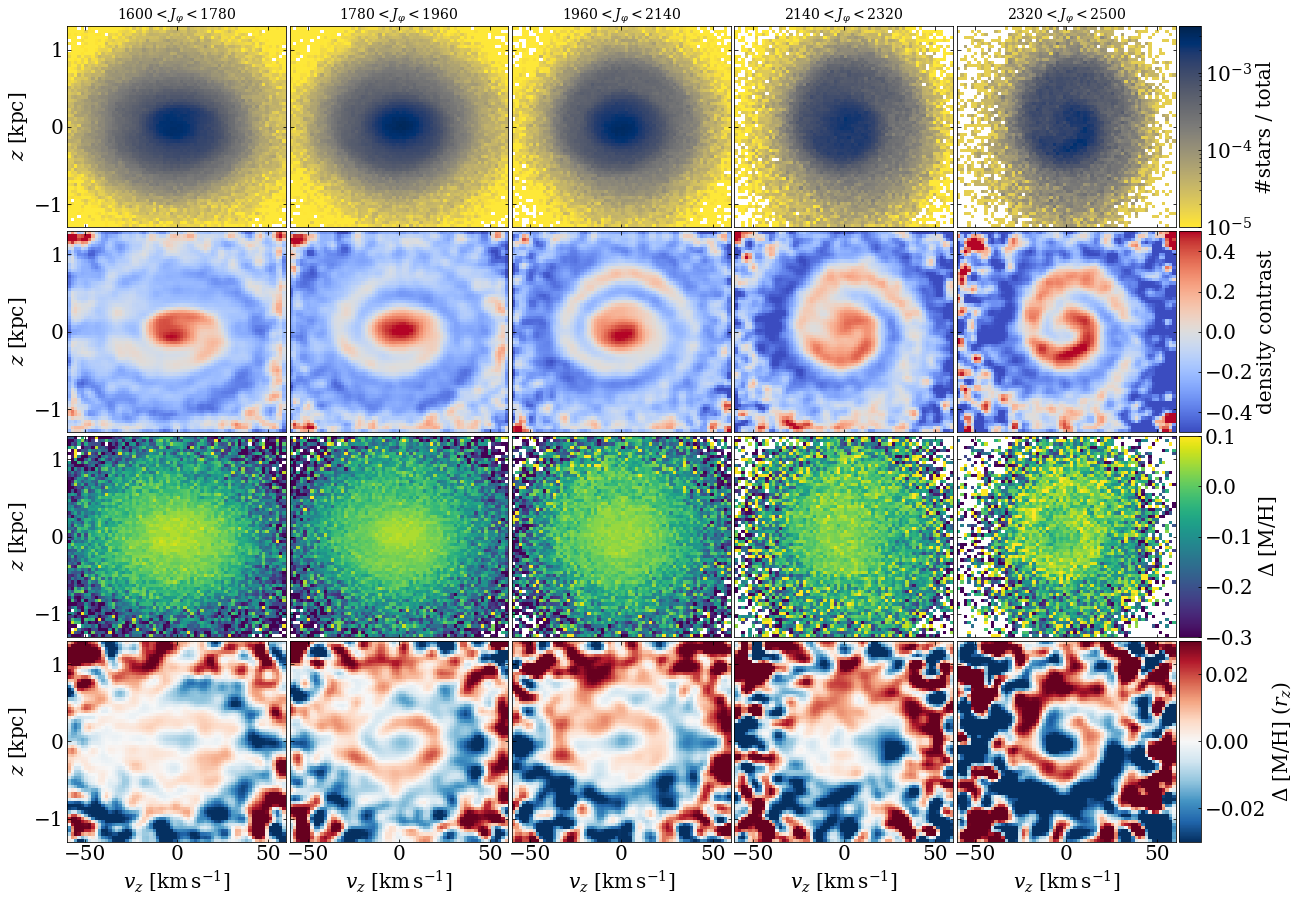}
    \caption{$z$--$v_z$ plane (position and velocity normal to the Galactic midplane) of the Gaia DR3 Medium Quality sample as defined in \cite{gaia_DR3_chemical_cartography}. Top row:  normalized star count in bins of angular momentum $J_\varphi$; the clockwise spiral structure is the Gaia Snail. There are 2\,477\,534 stars in this sample. Units of angular momentum are \kpckms.
    Second row: fractional density contrast $\Delta f /\bar{f}$ obtained by taking the residuals between the top row and a smoothed version of it. 
    Third row: $\Delta\mbox{[M/H]}\equiv \mbox{[M/H]}-\langle\mbox{[M/H]}\rangle$ where $\langle\mbox{[M/H]}\rangle$ is the mean metallicity in each panel as determined by Gaia. Bottom row: $\Delta\mbox{[M/H]}(r_z) \equiv \mbox{[M/H]} - \langle \mbox{[M/H]}\rangle(r_z)$, i.e., [M/H] residuals after subtracting the trend between mean metallicity and orbit (using orbit proxy $r_z$, which is closely related to the vertical action $J_z$, see description in Section \ref{section:OTI}), revealing a stronger Snail signal. The spiral structure seen in the third and fourth rows, which we call the Iron Snail, roughly follows the shape of the Gaia Snail in the first and second rows. In particular, regions of the phase-space plane with higher metallicity also tend to be denser. The shapes of the Snails in density and metallicity are not exactly the same because the gradients in density and metallicity are different in the original unperturbed system (see eq.\ \ref{eq:iron}).}
    \label{fig:Fig 1}
\end{figure*}

Galaxies grow hierarchically, via mergers with other galaxies, or by accreting gas \citep[e.g.,][]{rees_ostriker_1977,white_rees_1978,mo_mao_white_1998}. These mergers and close passages of satellite galaxies  continually leave dynamical imprints on their host galaxies. Therefore, galaxies are never in exact dynamical equilibrium: they are continually perturbed. These perturbations are manifest in the Milky Way \citep[e.g.,][]{widrow_2012, schoenrich_dehnen_2018}, especially in the exquisite data sets produced by the Gaia spacecraft.

After a perturbation has occurred to a stellar system, it phase mixes or Landau damps towards a new steady state. This new state is determined by the stars' new orbits, summarized by their new energies, or actions, in the new gravitational potential. These phase-mixing processes often manifest as shells around elliptical galaxies \citep{hq1988}, stellar streams \citep{helmi_white_1999, klement_2010, helmi_2020} and spirals or more complicated structures in phase space \citep{tremaine_1999}. The most beautiful example in the Milky Way is the Gaia Snail \citep[see][and Fig.\ \ref{fig:Fig 1}]{antoja_2018}, a spiral structure seen in the density  of stars in the solar neighborhood's vertical phase space (position and velocity normal to the Galactic midplane).

The Gaia Snail has been proposed to arise from various perturbations, such as the recent disk crossing of the Sagittarius dwarf galaxy \citep[e.g.,][]{antoja_2018, laporte_2019_sgrfootprint, bland-hawthorn_2019, bland-hawthorn_tepper-garcia_2021}, a dark-matter wake in the Milky Way halo \citep[e.g.,][]{grand_2022}, buckling of the Milky Way bar \citep{khoperskov_2019}, slowing of the bar \citep{li_2023}, frequent weak perturbations or gravitational ``noise'' \citep{tre22}, misalignment between the stellar disk and other structure (e.g., gas, inner dark-matter distribution) or, very plausibly, a combination of several of the above, making any inference about the origin of the Snail  challenging \citep{garcia_conde2024}.

If there is to be any hope of disentangling these hypotheses, the Snail must be characterized quantitatively and compared to the predictions \citep[e.g.,][]{antoja_2023, darragh-ford_2023}. In \cite{frankel_2023}, we quantified the Snail by constructing a simple model for the phase-space density, characterized by a perturbation amplitude, a perturbation ``age'' (or degree of winding) and an initial phase, all of which vary with the angular momentum or mean Galactocentric radius of the stars. We fitted these three parameters to the phase-space density as determined by the Gaia data, accounting for the selection effects due to the magnitude limits in our sample. 
However, other spatial selection effects such as spatially varying dust absorption have a direct imprint on the density in the data set \citep{frankel_2023}, limiting what we can learn from the Gaia data, especially at larger distances from the Sun.

In this paper, we focus instead on the signature of non-equilibrium dynamics on ``labels'', by which we mean time-independent properties intrinsic to the stars such as mass, metallicity or other abundances, etc. (see \S \ref{section:scott_math} for a more complete definition). We show that dynamical perturbations should \emph{generically} create structures in all labels so long as there are phase-space gradients in their mean unperturbed values. We use the Gaia Snail as an example of this process. We specifically use the metallicity, or iron abundance (hence the term Iron Snail\footnote{In the rest of this paper, we use the term ``Iron Snail'' generally, even when referring to other labels. We clarify which labels we use when relevant.}), together with a few additional labels, as illustrations of this argument. We then quantify the perturbation parameters of the Iron Snail (dynamical age or winding, amplitude, phase) as a function of angular momentum, and compare these to the parameters estimated from the phase-space density.

There are at least two advantages to tracking non-equilibrium dynamics using labels rather than phase-space density: (i) 
Spatial selection functions such as sky position-dependent magnitude limits and dust absorption have a much smaller effect on labels than on density (e.g., seeing fewer stars in some regions of the sky affects their density but not their mean metallicity, beyond increasing the uncertainty on the mean). (ii) If there is more than one label per star, we can in principle quantify the perturbation in additional dimensions, for example, if the Snail contains both vertical and radial perturbations \citep[as shown by all simulations so far and hinted to by observations, e.g.,][]{carr_2022, hunt_2021, bland-hawthorn_tepper-garcia_2021, hunt_2024_radial_phase_spirals}.

Stellar labels have long been used to trace the chemo-dynamical history of the Milky Way. Because labels are invariant in time, and reflect properties of the birth environment of stars, they can be used to trace the formation and subsequent history of our Galaxy. This line of analysis is known as chemical tagging \citep{freeman_bland-hawthorn_2002}, and is closely related to the work in this paper: stellar labels retain information that pre-dates dynamical perturbations.

The use of labels to track non-equilibrium dynamics is not a new idea either. 
Stellar labels have been used to trace the dynamics of stars in the Milky Way disk \citep{price-whelan_2024, horta_2024,price-whelan_2021_OTI} and part of this analysis is based on the results from these works.
\cite{gaia_DR3_chemical_cartography} and then \cite{alinder_2023} showed evidence of the Snail when the vertical phase-space plane is color-coded by mean [M/H]. We reproduce this result \cite[][Fig.\ 20]{gaia_DR3_chemical_cartography} in the third row of Figure \ref{fig:Fig 1}, and, by subtracting the overall trend in metallicity versus the amplitude of vertical excursions (i.e., the vertical action $J_z$), we show that the data reveal a convincing spiral signal in the residuals (bottom row of Figure~\ref{fig:Fig 1}). Previous works have shown that the Snail also appears when color-coding the vertical phase-space plane by the azimuthal velocity $v_\varphi$ \citep{antoja_2018}; here  $v_\varphi$ is a proxy for angular momentum, which is an integral of motion and therefore a suitable label. 

\section{The Emergence of Iron Snails: Data and Interpretation}

\begin{figure}
    \centering
    \includegraphics[width=0.5\textwidth]{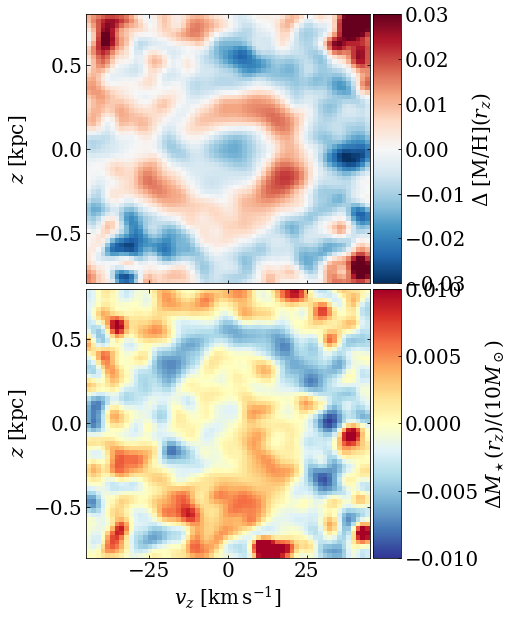}
    \caption{Top panel: $z$--$v_z$ plane for the mean metallicity in the Gaia DR3 Medium Quality sample. The colors represent residuals after subtracting the smoothed mean metallicity field. The figure is similar to the second panel from the left in the bottom row of Figure \ref{fig:Fig 1} and contains $1.57\times 10^6$ stars. Bottom panel: a similar plot for 68\,241 stars having asteroseismic masses from TESS. The angular-momentum range in each panel is $1700 \leq J_\varphi \leq 2000\,\mathrm{kpc\,km\,s^{-1}}$. The Snail signal is qualitatively similar in both labels, with regions containing higher mass (so on average younger) stars coinciding with regions containing metal-rich stars.}
    \label{fig:Fig 2}
\end{figure}

In this section, we qualitatively explore the vertical phase space (distance $z$ and velocity $v_z$ normal to the Galactic midplane) of stars in three different datasets, color-coding by the means of different labels in phase-space pixels, $\langle {\stag} \rangle (z, v_z)$, where $\stag$ can be metallicity [M/H], asteroseismic mass, stellar age, etc. (i) A sample of 2\,477\,534 stars close to the one revealing the [M/H] spirals in Gaia DR3 \citep[][Fig.\ 20]{gaia_DR3_chemical_cartography}, already shown briefly in Figure \ref{fig:Fig 1}. (ii) A sample of 68\,241 stars with asteroseismic masses from the TESS spacecraft \citep{hon_2021}. (iii) A sample of 29\,485 stars from the ground-based APOGEE DR17 survey, which provides several abundance measures ([Fe/H], [Mg/Fe], [C/N]) as well as estimates of stellar ages \citep{sdss_dr17_2022, leung_bovy_2019}. 

\subsection{Emergence of Iron (or label) Snails in Various Datasets}
\begin{figure*}
    \centering
    \includegraphics[width=\textwidth]{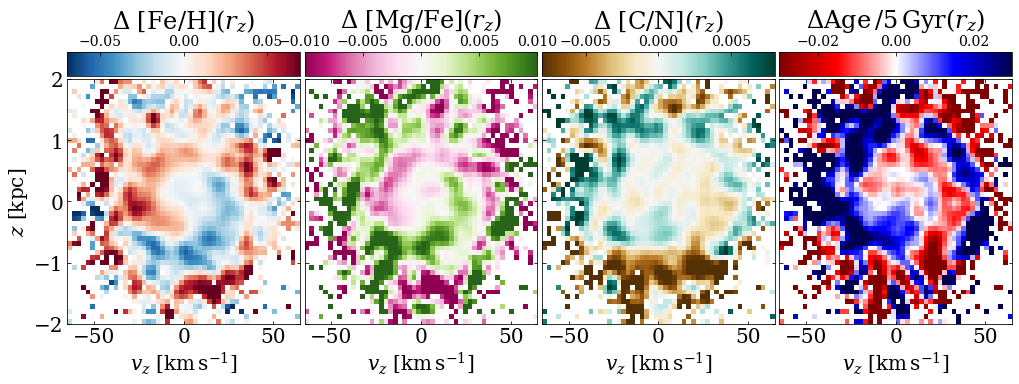}
    \caption{$z-v_z$ plane color-coded by four label residuals as in the bottom row of Figure 1: [Fe/H], [Mg/Fe], [C/N], and age, from the APOGEE DR17 dataset. The 29\,485 stars in this sub-sample have angular momentum $2200 \leq J_\varphi \leq 3800 \,\kpckms$. The Snail is qualitatively similar in all labels.}
    \label{fig:Fig 3}
\end{figure*}

Here, we specify how we use the three samples (Gaia, TESS and APOGEE) to produce Figures \ref{fig:Fig 1}--\ref{fig:Fig 3}. We start with the Gaia DR3 dataset, select stars as in the Medium Quality Sample of \cite{gaia_DR3_chemical_cartography}, with parallax $\varpi$ and parallax error $\sigma_\varpi$ given by  $1/\varpi < 3$ kpc, $\varpi/\sigma_\varpi > 5$, and a simple cut for brighter and redder stars with $M_G < 4$ and $\mbox{BP}-\mbox{RP} > 1$, and we are left with 2 477 534 stars. Following \cite{gaia_DR3_chemical_cartography}, we use the metallicity estimates produced by the Gaia \textit{GSP-Spec} module (i.e., the data column named 
{\tt mh\_gspspec}) 
which uses the information from the higher resolution RVS spectra \citep{recio-blanco_2023_gaia_dr3}. We correct the raw metallicity values from the catalogs with log $g$-dependent offsets as prescribed in \cite{recio-blanco_2023_gaia_dr3}; these were determined by these authors through comparisons with spectroscopic estimates of surface abundances from other sources.

We transform to Galactocentric coordinates assuming the distance to the Galactic centre and the height of the Sun to be respectively $R_\odot = 8.23 $ kpc \citep{leung_2022} and $z_\odot = 20.8$ pc \citep{bennett_bovy_2019}, a solar motion with respect to the Local Standard of Rest of $v_\odot=(11.1, 12.24, 7.25)\,\mbox{km s}^{-1}$\citep{schoenrich_2010} and a circular velocity at the solar radius of $220\,\mathrm{km\,s^{-1}}$ as implemented in \texttt{Galpy} \citep{bovy_2015_galpy}.
We then split the stars into five angular-momentum bins between 1600 \kpckms\ and 2500 \kpckms. Figure \ref{fig:Fig 1} shows the Snail as it appears using different labels at different angular momenta, as follows. 

The top row is the (raw) number of stars in phase-space pixels, i.e., the number of stars in our catalog, without any kind of selection correction or re-weighting.
The second row shows the fractional density contrast of the observed phase-space distribution function (i.e., from the top row) relative to a smoothed version of it, and reveals the details of the Gaia Snail.
The third row depicts the mean metallicity of the stars in phase-space pixels, as determined by Gaia and corrected according to Table 3 in \cite{recio-blanco_2023_gaia_dr3} (\verb|mh_gspspec|) after the average metallicity in a given panel is subtracted.\footnote{Note that \cite{gaia_DR3_chemical_cartography} additionally remove the average metallicity in 0.1 kpc radial bins, whereas here we only subtract the mean metallicity in each panel.} The Snail is not as visible as one would like because because the signal is obscured by the label gradient in the unperturbed system: the center of the $z$--$v_z$ plane is dominated by metal-rich stars while the outskirts are dominated by metal-poor stars.
The third row addresses this issue by showing the [M/H] residuals after removing the [M/H] gradients from a best-fit Orbital Torus Imaging model (OTI, described in further detail in Section \ref{section:OTI}). With this refinement, the Iron Snail becomes obvious in most of the panels.

We next turn to the sample from TESS \citep{hon_2021}, and produce again residuals between a label (now asteroseismic mass) and a best-fit model for how the mean label depends on orbits in the $z$--$v_z$ phase space (see Section \ref{section:model} for the math). We use Gaia's astrometric values and invert the Gaia parallaxes of stars in the TESS sample to get approximate distances. We convert the coordinates to the Galactocentric reference frame in the same way as for the Gaia sample in the previous section, with the same solar position, velocity, and Local Standard of Rest.
We select stars with angular momenta $1700 \leq J_\varphi \leq 2000 \mathrm{\,kpc\,km\,s^{-1}}$, to focus on an angular-momentum bin where the Snail is already well visible in Gaia metallicities in Figure \ref{fig:Fig 1}. This bin is centered on the second column in Figure \ref{fig:Fig 1}, but is slightly wider to keep a larger number of stars in the sample (68\,241) to reduce the noise. The median distance of stars in this reduced TESS sample is about 0.9 kpc. The bottom panel of Figure \ref{fig:Fig 2} illustrates the resulting residuals; the top panel indicates the residuals in metallicity from the Gaia sample described in Figure \ref{fig:Fig 1}. The two labels reveal a similar spiral feature. Metal-rich regions of phase space seem to contain, on average, more massive stars too (as expected, since massive stars are younger and young stars are more metal-rich). 

Finally, we explore a few more labels from the APOGEE DR17 sample \citep{sdss_dr17_2022}. APOGEE is a pencil-beam spectroscopic survey that measured precise abundances (at the $0.05$ dex level) from near-infrared spectra. The spatial selection is complex, in part because of the pencil beams, and therefore this sample is not well suited to study the Gaia Snail in density space. But as long as the mean labels (abundances, ages, etc.) are unaffected by the spatial selection, the Iron Snail should be unaffected. In Figure \ref{fig:Fig 3}, we show the residuals from a best-fit mean label for the abundances [Fe/H], [Mg/Fe], [C/N], and for stellar age from \cite{leung_bovy_2019}. This sub-sample of stars was selected to (i) be in the Main Survey of APOGEE, (ii) have an in-plane projected distance less than 5 kpc from the Sun, (iii) have $3500 < T_\mathrm{eff} < 6500$ K and $1.5 < \mathrm{log}g < 3.4$. The resulting sample spans several kpc around the Sun, and has a mean distance from the Sun of about 3 kpc. The Snail-like features are present and similar in all labels, although they are more visible in some labels ([Fe/H], [Mg/Fe]) than others ([C/N]).

\subsection{Qualitative Observations}
From these analyses, we can draw a few observations.
\begin{enumerate}
    \item The Snail has qualitatively the same shape in all labels we have examined, but differs in the details.
    
    \item The denser parts of the Snail coincide with the metal richer ones (Figure \ref{fig:Fig 1}). This is consistent with the observation that the density gradient and the metallicity gradient in the equilibrium system follow each other (this is true both radially \emph{and} vertically).
    
    \item The metal richer regions of the Snail coincide with those having larger asteroseismic masses (Figure\ \ref{fig:Fig 2}) and younger ages (Figure  \ref{fig:Fig 3}). This is consistent with the observation that high-mass stars tend to be young and metal-rich. 
    
    \item On average, massive stars are younger than low-mass stars due to their shorter life time. In the Milky Way, there is a strong positive vertical age gradient (often explained with vertical dynamical heating, or upside-down disk formation), and a weak negative radial age gradient (often explained with an inside-out growth of the disk). The presence of the Snail in the age label confirms that the vertical gradients are important, i.e., that the Snail appears because the vertical frequency depends on vertical action $\Omega_z = \Omega_z(J_z)$, not mainly because it depends on angular momentum or azimuthal action $\Omega_z = \Omega_z(J_\varphi)$\footnote{The general case is $\Omega_z = \Omega_z(J_\varphi, J_z)$.}. In other words, the vertical age gradient in the Galactic disk is much stronger than the radial one: $| \partial \langle \tau \rangle /\partial J_\varphi| \ll |\partial \langle \tau \rangle /\partial J_z| $, i.e., $ \langle \tau \rangle$ depends on $\Omega_z$ mainly via $J_z$, not via $J_\varphi$ and therefore shows that the vertical projection of the perturbation is important, since a purely planar perturbation would not produce a phase-space structure in the age label. 
\end{enumerate}
This qualitative picture hints at the link between dynamical perturbations and abundance structures. Because stars retain their elemental abundances from birth, a dynamical perturbation that changes the orbits of stars and produces kinematic structure will also produce a structure in abundances (or any other approximately conserved quantity in the stars). The next section lays out the quantitative derivation of this argument.

\subsection{Quantitative Interpretation: The General Case\label{section:scott_math}}

\begin{figure*}
    \centering
    \includegraphics[width=\textwidth]{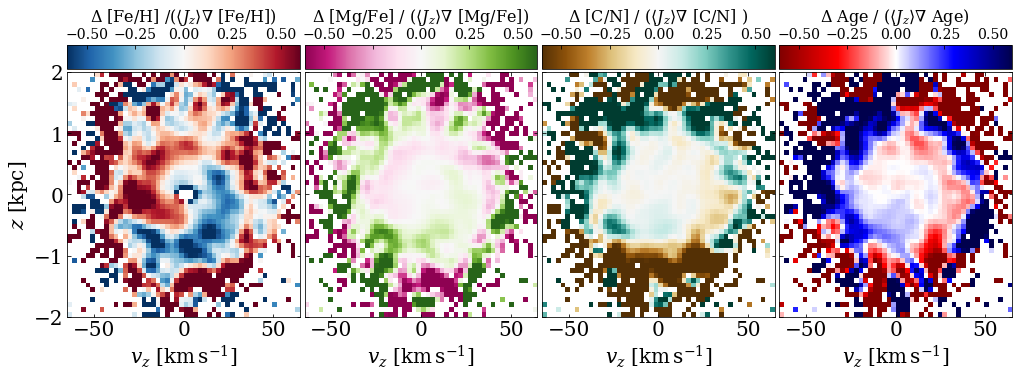}
    \caption{$z$--$v_z$ plane, color-coded by normalized residuals for the APOGEE DR17 dataset. The plots show the right-hand side of equation (\ref{eq:yyy}). In the captions we denote the label gradient $\nabla \stag \equiv \partial \bar{\stag}/\partial J_z$. The range of angular momentum is $2200 \leq J_\varphi \leq 3800 \,\kpckms$. The Snail signal (color code) is broadly similar in all labels. We took $\langle J_z \rangle = 10 \,\kpckms$. The signal strength varies between about 0.2 and 0.6 depending on the position in phase space and the label. The central white area in the center of the left panel arises because the best-fit vertical metallicity gradient becomes too close to zero in the mid-plane.}
    \label{fig:Fig_normalized_res_eps}
\end{figure*}

The emergence of the Snail in different labels can be explained by considering an initially unperturbed distribution of stars that presents label gradients, and perturbing it. Since these labels are invariants, the perturbation produces label phase spirals that wind up in the same way that the density does.

Let $(\bq,\bp)$ be vectors representing a set of canonical coordinates and momenta in a stellar system, and $f(\bq,\bp,t)$ the distribution function of stars in phase space, i.e., $f(\bq,\bp,t)d\bq d\bp$ is the number of stars at time $t$ in the phase-space volume element $(\bq,\bp)\to(\bq+d\bq,\bp+d\bp)$\footnote{In our notation, if $q,p$ are bold ($\bq, \bp$) or have subscripts ($q_0, q_z, p_0, p_z$), then they represent phase-space coordinates; else $p$ represents a normalized probability density function while $f$ represents an un-normalized distribution function.}. Then if stars are neither created nor destroyed the distribution function satisfies the collisionless Boltzmann equation,
\begin{equation}
    \frac{Df}{Dt}=\frac{\p f}{\p t}+[f,H]=0,
    \label{eq:cbe}
\end{equation}
where $Df/Dt$ is the Lagrangian derivative of $f$ in phase space, $H(\bq,\bp,t)$ is the Hamiltonian governing the motion of the stars, and the Poisson bracket of two phase-space functions $a(\bq,\bp,t)$, $b(\bq,\bp,t)$ is
\begin{equation}
    [a,b]\equiv \frac{\p a}{\p\bq}\cdot\frac{\p b}{\p\bp} -\frac{\p b}{\p\bq}\cdot\frac{\p a}{\p\bp}.
\end{equation}

We define a label to be a conserved quantity associated with a star that does not depend on its phase-space evolution or have any explicit time dependence. Example labels could include the stellar mass (if mass loss and accretion are negligible), color (if stellar evolution and time-variable reddening are negligible), metallicity or more generally relative element abundances, the strength of a particular spectral feature, the angular momentum in an axisymmetric potential, etc. If $\stag$ is a label, then so is any function $g(\stag)$.

Let us define a generalized distribution function $f(\bq,\bp,\stag, t)$ 
such that  $f(\bq,\bp,\stag,t)d\bq d\bp d\stag$ is the number of stars at time $t$ in the phase-space volume element $(\bq,\bp)\to(\bq+d\bq,\bp+d\bp)$ with labels in the range $\stag\to \stag+d\stag$. Then $f(\bq,\bp,\stag,t)$ satisfies the collisionless Boltzmann equation (\ref{eq:cbe}) as well. 

The mean label $\bar{\stag}$ in a phase-space element is
\begin{equation}
    \bar{\stag} (\bq,\bp,t)\equiv \frac{\int d\stag\,\stag f(\bq,\bp,\stag, t)}{\int d\stag\,f(\bq,\bp,\stag,t)}.
    \label{eq:zbar}
\end{equation}
It is straightforward to show that the mean label satisfies the collisionless Boltzmann equation. Many other functionals of the generalized distribution function also satisfy the collisionless Boltzmann equation; for example
\begin{equation}
    w(\bq,\bp,t)\equiv \frac{\int d\stag\,G_1[\stag,f(\bq,\bp,\stag,t)]}{\int d\stag\,G_2[\stag,f(\bq,\bp,\stag,t)]},
    \label{eq:mg}
\end{equation}
with $G_1$ and $G_2$ arbitrary differentiable functions.
The average $\bar{\stag}$ in equation (\ref{eq:zbar}) is an example of such a functional. 


These simple arguments suggest that if a Snail is present in the distribution function $f(\bq,\bp,t)$ then one should also be present in the mean label $\bar{\stag}(\bq,\bp,t)$, e.g., in the mean metallicity $\langle [M/H] \rangle (\bq,\bp,t)$.
The relative strengths of the Snails in these two functions depend on the gradients in the functions in the original unperturbed state; for example, if there is initially no gradient in the mean metallicity then no Snail will develop in $\bar{\stag}(\bq,\bp,t)$. 

To explore the relation between the Snail (or other non-equilibrium structures) in the distribution function and in the mean metallicity, we examine a stellar system with one degree of freedom. The prototypical system of this kind in the Galactic context is an infinite disk with slab symmetry, in which the coordinate and momentum can be chosen to be the distance and velocity relative to the midplane, $(z,v_z)$. Such systems provide the simplest models in which we can explore the formation and evolution of a Snail. We assume that initially (time $t<0$) the gravitational potential, the distribution function, and the phase-space distribution of the mean label are all in a steady state. We write the latter two functions as $f_0(z,v_z)$ and $\bar{\stag}_0(z,v_z)$. According to the Jeans theorem \citep{bt08}, any steady-state solution of the collisionless Boltzmann equation depends on the phase-space coordinates $(z,v_z)$ only through integrals of motion. Therefore $\bar{\stag}_0(z,v_z)$ must be an integral of motion, and the distribution function can be written in the form $f_0(z,v_z)=F[\bar{\stag}_0(z,v_z)]$ for some function $F \geq 0$.

Now let us introduce an external perturbing potential, starting at time $t=0$. The distribution function and mean metallicity now become functions of time, $f(z,v_z,t)$ and $\bar{\stag}(z,v_z,t)$ for $t>0$, which satisfy the collisionless Boltzmann equation. Now consider the phase-space function $F[\bar{\stag}(z,v_z,t)]$. As we have argued, for $t<0$ this function is identical to the distribution function, and at all times they both satisfy the collisionless Boltzmann equation. Therefore $f(z,v_z,t)=F[\bar{\stag}(z,v_z,t)]$ at all times $t>0$. In other words, at any given time a Snail or other non-equilibrium feature must be identical in the distribution function $f(z,v_z,t)$ and the nonlinear function of the mean metallicity $F[\bar{\stag}(z,v_z,t)]$. 

These relations are simplified when the perturbation to the initial equilibrium distribution function is small. Write $f(z,v_z,t)=f_0(z,v_z)+\epsilon \delta f(z,v_z,t)$ and $\bar{\stag}(z,v_z,t)=\bar{\stag}_0(z,v_z)+\epsilon \delta\bar{\stag}(z,v_z,t)$, with $\epsilon \ll 1$. Expanding the relation $f(z,v_z,t)=F[\bar{\stag}(z,v_z,t)]$ to first order in $\epsilon$ we find
\begin{equation}
    \delta f(z,v_z,t)= F'[\bar{\stag}_0(z,v_z)]\delta\bar{\stag}(z,v_z,t).
\end{equation}
If the initial distribution function and mean label are written as functions of some other integral $I(z,v_z)$ (for example the energy or action), $f_0(I)$ and $\bar{\stag}_0(I)$, then this expression can be rewritten as
\begin{equation}
    \frac{\delta f(z,v_z,t)}{f_0(z,v_z)}= \frac{ d\ln[ f_0(I)]/dI}{d\bar{\stag}_0(I)/dI}\delta\bar{\stag}(z,v_z,t).
    \label{eq:iron}
\end{equation}
Since the perturbations are small, $f_0(I)$ and $\bar{\stag}_0(I)$ can be estimated using a smoothed form of the observed functions $f(z,v_z,t)$ and $\bar{\stag}(z,v_z,t)$. The simplest algorithm for smoothing is to write the functions in terms of  angle-action variables $(\theta_z,J_z)$, and then to average over the angle $\theta$. For the plots in this paper, we use the vertical action as integral of motion, $I = J_z$ and further simplify $f_0(J_z)$ by assuming it to be exponential, $f_0(J_z)\propto \exp(-J_z/\langle J_z\rangle)$. Then equation (\ref{eq:iron}) becomes
\begin{equation}
    \frac{\delta f(z,v_z,t)}{f_0(z,v_z)}= -\frac{ 1}{\langle J_z\rangle d\bar{\stag}_0(J_z)/dJ_z}\delta\bar{\stag}(z,v_z,t).
    \label{eq:yyy}
\end{equation}
Figure \ref{fig:Fig_normalized_res_eps} plots the right-hand side of this equation, and can be compared to the second row of Figure \ref{fig:Fig 1}, which plots the left-hand side. Both plots are consistent with the range $0.2 \lesssim \delta f/f_0 \lesssim 0.5$, which is a similar range to  previously measured density contrasts in the literature \citep[][although apple-to-apple comparions are limited]{frankel_2023, alinder_2023}.

Equations (\ref{eq:iron}) or (\ref{eq:yyy}) provide the expected relation between the Snail or other non-equilibrium features in the mean label and in the phase-space density when the stellar system has one degree of freedom. Assuming that the system has one degree of freedom is a reasonable first approximation when examining the dynamics of the Galactic disk in the direction normal to the Galactic midplane, since the gradients in the density and metallicity are much stronger in this direction than in the radial or azimuthal direction. 

In principle, these results can be extended to stellar systems with two or three degrees of freedom, provided that there are two or three labels that are well-measured and independent. However, in this paper we will restrict ourselves to motion with a single degree of freedom, which we take to be the direction normal to the Galactic midplane.

\section{Iron Snail Modeling: Empirical Orbit Labels and Perturbation\label{section:model}}

We now quantify the Iron Snail signal seen in APOGEE with a forward model.
We aim to determine the parameters of the perturbation that could lead to the Iron Snail (and similar structures in other abundances). We start by deriving empirical orbit labels, i.e., actions, angles and frequencies. 
We then construct a simple single-impact model that perturbs the orbits of stars in phase space, and let the perturbation wind up with time $t$. We constrain this model against the data and infer a perturbation amplitude and dynamical age.

\subsection{Empirical Orbit Labels \label{section:OTI}}
We use a version of the Orbital Torus Imaging (OTI) package described in \citet{price-whelan_2021_OTI,price-whelan_2024} to infer empirical vertical actions ($J_z$), frequencies ($\Omega_z$) and angles ($\theta_z$). This method does not derive these quantities from an assumed gravitational potential. Rather, it assumes that the stars are well mixed, so their mean abundances and other labels should be independent of their orbital phase (or angles). Therefore, contours of constant mean abundance in phase space should trace individual orbits. OTI presumes phase space can be mapped from configuration space $(z, v_z)$ to an angle-action proxy space ($\tilde{\theta}, r_z$), where the mean abundance would depend only on the action proxy $r_z$ and not on $\tilde{\theta}$. Once this mapping has been found, the orbital parameters can be worked out from the orbit shapes. Namely, finding the vertical action, period, and angle amounts to integrating phase space along lines of constant $r_z$, as in \cite{price-whelan_2024}, eqs.\ (24)--(26),
\begin{equation}
    \begin{split}
        J_z &= \frac{2}{\pi} \int_0^{\pi/2} d\tilde{\theta}\, v_z(\tilde{\theta}) \bigg|\frac{dz}{d\tilde{\theta}}\bigg| \\
        T_z &= 4 \int_0^{\pi/2} \frac{d\tilde{\theta}}{v_z(\tilde{\theta})} \bigg|\frac{dz}{d\tilde{\theta}}\bigg| \\
        \theta_z &= \frac{2\pi}{T_z}\int_0^{\tilde\theta^*} \frac{d\tilde{\theta}}{v_z(\tilde\theta)}  \bigg|\frac{dz}{d\tilde{\theta}}\bigg| 
    \end{split}
\end{equation}
with the vertical frequency $\Omega_z$ related to the vertical period $T_z$ by $\Omega_z = 2\pi/T_z$. In the last equation, $\tilde{\theta^*}$ is the angle as measured at the position of a star (rather than an integration variable). We refer the reader to \cite{price-whelan_2024} for additional details.

If using several labels at a time, for example as in Figures   \ref{fig:Fig 3} and \ref{fig:Fig_normalized_res_eps}, we fit these OTI models using all labels simultaneously, where each label-orbit relation is label-dependent, but the $z$--$v_z$ orbit relation must remain the same regardless of the label used. In other words, we assume there is a unique $r_z = \xi(z,v_z)$ relation at given $J_\varphi$, and that for $N$ labels \stag, there are $N$ relations $\bar{\stag} = g(r_z)$. The residuals left from these procedures should reveal any ongoing phase mixing structure because the model assumes a fully phase-mixed system. Residuals in [Fe/H], [Mg/Fe], [C/N] and age from \cite{leung_bovy_2019} are shown in Figure \ref{fig:Fig 3}. 

\subsection{Modeling the Iron Snail}

We model the mean abundance, or any label $\bar{\stag}(\J_0)$,~as arising from an original relation between the label and the unperturbed actions. After a dynamical perturbation, the orbits change and acquire new actions and the perturbation winds up, but the stellar labels don't change: they remain simple functions of the old orbits (the unperturbed actions). 

The equations in this section are close to those in Section \ref{section:scott_math}, except that for fitting purposes, we now work with normalized probability densities $p$ rather than un-normalized phase-space density $f$. We start by perturbing the orbits of stars from an initial state at time $t=0$, and work out how this would affect the mean label as a function of phase-space position at any later time $t$. The model parameters to be fit are the perturbation amplitude, dynamical age and initial phase. 

The mean label as a function of the angle-action coordinates of the current orbit is
\begin{equation}
\begin{split}
    \langle \stag \rangle (\J,\ang,t) &= \int \stag\, p(\stag | \J, \ang, t) d\stag \\
                           &= \iint \stag\, p(\stag | \J_0)  d\stag p(\J_0 | \J, \ang, t) d\J_0,
\end{split}
\end{equation}
where we inserted the unperturbed actions $\J_0$ in the second line. If we assume that (i) the system is collisionless (i.e., there is no diffusion of actions) and (ii) the perturbation is impulsive, then we can write 
\begin{equation}
     \prob(\J_0 | \J, \ang, t) = \delta(\J_0 - \J_{0,{\rm func}}(\J, \ang, t))
\end{equation}
where $\J_{0,{\rm func}}(\J, \ang, t)$ is to be specified by our model for the impulse, described below (it is found by solving equations \ref{eq:q0p0} and \ref{eq:qp}).
The integral becomes
\begin{equation} \label{eq:mean_label}
\begin{split}
    \langle \stag \rangle(\J,\ang,t) &=\int \bar{\stag}(\J_0) p(\J_0 | \J, \ang, t) d^3\J_0\\
                           &=  \bar{\stag}(\J_{0,{\rm func}}(\J,\ang,t)).
\end{split}
\end{equation}
We need to specify the relation between unperturbed action and label. In the following, we take this relation to be the unperturbed $\bar{\stag}(\J_{0}) = \bar{\stag}(\J)$ as determined by OTI, which is correct (only) to first order in the perturbation.

For application to the simple vertical phase-space structure, we now work in only one dimension, so  $(\J,\ang)$ simplifies to $(J_z, \theta_z)$.
We specify the impulsive perturbation in canonical Cartesian coordinates: before the perturbation the Cartesian coordinates are related to the angle and action by 
\begin{equation}\label{eq:q0p0}
    \begin{split}
        q_0 &= \sqrt{2J_0}\sin(\theta_0) \\
        p_0 &= \sqrt{2J_0}\cos(\theta_0),
    \end{split}
\end{equation}
and afterward the relation is 
\begin{equation} \label{eq:qp}
    \begin{split}
        q_z &= q_0 + \Delta q_z = \sqrt{2J_z}\sin(\theta_z) \\
        p_z &= p_0 + \Delta p_z = \sqrt{2J_z}\cos(\theta_z)
    \end{split}
\end{equation}
where $\Delta q_z$ and $\Delta p_z$ are the parameters of the kick, assumed independent of position in phase space.
The amplitude of the impulse in our coordinates of interest, the orbital action, is $\delta J = \frac{1}{2}(\Delta q_z^2 + \Delta p_z ^2)$ and its phase $\varphi = \mbox{arctan2}(\Delta q_z, \Delta p_z)$. After the perturbation, we assume that the stars subsequently phase mix as $\theta_z = \Omega_z(J_z) t+\mbox{const}$ with their new actions and frequencies which remain constant.

Equation (\ref{eq:mean_label}) shows that we can, in principle, predict the mean label as a function of phase-space position. It is independent of spatial selection effects as long as the selection does not depend on quantities that correlate with abundances. This means that if a survey selects more stars in some regions of the sky than others, as the pencil-beam survey APOGEE does, the \emph{observed} mean label will not be affected (beyond varying Poisson noise), even if the \emph{observed} phase-space density is affected. Here, we want to exploit this property of the mean abundances.

We can then relate the \emph{average} strength of the dynamical perturbation to the phase-space density perturbation $\delta f/f_0$ as in Section \ref{section:scott_math}. 
To first order in the kick amplitude, the perturbation to the distribution from a kick that occurs at a time $t=0$ can be written as
\begin{equation}
    p(J_z, \theta_z) = \frac{1}{2\pi}p_\mathrm{eq}(J_z)\left[1 + A_J(J_z)\cos(\theta_z-\Omega_z t -\varphi) \right],
\end{equation}
where $p_\mathrm{eq}(J)$ is the unperturbed probability distribution of the vertical action (i.e., the action distribution before the kick), and
\begin{equation}
    A_J(J_z) = - 2\sqrt{J\delta J} \frac{d\ln p_\mathrm{eq}(J)}{dJ}.
\end{equation}
For the purpose of (i) simplicity and (ii) quantitative comparisons with our previous work in \cite{frankel_2023}, we want to re-write the distribution function with an amplitude $A$ that is constant, rather than a function of vertical action. We can average the amplitude of the perturbation as a function of $J_z$ to get $p(J_z, \theta_z) \simeq \frac{1}{2\pi}p_\mathrm{eq}(J_z)[1 +  A  \cos(\theta_z -\Omega_z t - \varphi)]$, with 
\begin{equation}
    A \equiv \langle A_J \rangle = \int p_\mathrm{eq}(J_z) A_J(J_z) dJ_z = \sqrt{ \frac{\pi \delta J}{\langle J_z\rangle}}
\end{equation} 
in the case where $p_\mathrm{eq}(J_z)$ is an exponential distribution with scale $\langle J_z \rangle $.  With this approximation, we can write

\begin{equation} \label{eq:perturbation_amplitude}
   \frac{\delta f}{f_0}  =  A \cos(\theta_z - \Omega_zt - \varphi) = \sqrt{\frac{\pi \delta J}{\langle J_z \rangle}}\cos(\theta_z - \Omega_zt - \varphi).
\end{equation}
Note that the label perturbation $\delta \bar{\stag}$ will have a similar form, see equations (\ref{eq:iron}) and (\ref{eq:yyy}). 
With this expression, we can relate our best-fit parameters $(\Delta q_z, \Delta p_z, t)$ to a relative perturbation of the phase-space density, which has been measured for the Gaia Snail in previous literature.

\vspace{6mm}
\subsection{Fitting the Iron Snail}
We model the distribution of label \stag~as a Gaussian distribution $\mathcal{N}$ with mean equal to the mean label value $\bar{\stag}(J_0)$  and standard deviation $\sigma_\stag$,
\begin{equation}
    p(\stag|z,v_z,\params) = \mathcal{N}(\stag- \bar{\stag}({J}_0({J_z},{\theta},\params)),\sigma_\stag),
\end{equation}
with parameters \params~setting the amplitude of the perturbation $A$, its dynamical `age' $t$, and initial phase. 
We can write a likelihood function assuming all the label measurements are independent, 
\begin{equation}
    \mathcal{L}(\params; \{\stag, z, v_z\}) = \prod_{i=1}^{N_\star}
    p(\stag_i|z_i,v_{zi},\params).
\end{equation}
We maximize this function and thereby obtain the best-fit parameters $\hat{\params}$. We estimate the uncertainties on the best-fit parameters by repeating the procedure 100 times on bootstrap samples from the same dataset (sampling with replacement). 

In practice, we apply this procedure using the [Mg/Fe] abundance because this label yields the most robust results (we defer fitting other element abundances to future work). We fix $\sigma_\stag = 0.3$ dex\footnote{We verified that the value of $\sigma_\stag$ does not influence the results by reproducing the results within their uncertainties with $\sigma_\stag = 0.03$ dex.}.

The resulting perturbation amplitude and age are shown in purple in Figure \ref{fig:amp_age_fit}; in deriving this figure we assumed a mean vertical action of $\langle J_z \rangle = 10 \,\kpckms$. Where the best-fit amplitude is low ($\lesssim 0.1$, i.e., at $J_\varphi < 1800\, \kpckms$), we recommend the reader not to interpret the values of the parameters because there is no clear Snail signal there, but the model can still infer a physically meaningless dynamical age. We left these ages visible in Figure \ref{fig:amp_age_fit}, but as dashed lines at lower opacity.

We performed several tests to confirm that our procedure was working correctly. (i) We started the optimizer from 100 different starting points and chose the best fit that has the maximum likelihood value to ensure that we were not stranded in a local maximum. (ii) We replaced the [Mg/Fe] values of the observed data by those we would expect from a known mock impulse (i.e., we `re-paint' the stars in the observed data), and checked that we can recover the true amplitude and perturbation age. (iii) We performed a full test on mock data where we generate both perturbed orbits and the labels associated with them (instead of keeping the kinematics of the observed data as is done in (ii)). There, the orbits are also perturbed according to an impulse, and the fitting procedure recovers the right parameters. 

Figure \ref{fig:enter-label} illustrates a comparison between the data and the best-fit model (plus noise) in the Zebra diagram (angle-frequency plane). In this plane, phase-wrapped features should be straight lines. The figure shows that where the signal amplitude is high, the best-fit model finds parameters that capture the slope (dynamical age) and phase reasonably well. Where the signal amplitude is low, the best-fit signal has non-interpretable parameters that mostly fit noise or a signal not well described by a phase-mixed perturbation. We note that in the angular-momentum range 1725--1800$\,\kpckms$ of Figure  \ref{fig:enter-label}, we could potentially see a 2-armed ($m=2$) perturbation (two parallel zebra lines in the same panel, separated by an angle $\pi$, instead of just one zebra line), which could be related to the 2-armed spiral identified in the inner disk by \citet{hunt_2022}. 

In principle, provided there are suitable non-degenerate gradients between labels and actions, we could fit a 3-dimensional perturbation (in all three actions). Here, we have only shown a proof of concept in the one-dimensional, vertical case and defer the combination of in-plane and out-of-plane perturbation to a further study: in-plane motion is complex, with several in-plane perturbations from the bar and spiral arms, and deserves to be investigated in greater depth than our ambition for this paper.

\subsection{Best Fit: The Gaia Snail vs.\ The Iron Snail}

\begin{figure}
    \centering
    \includegraphics[width=\columnwidth]{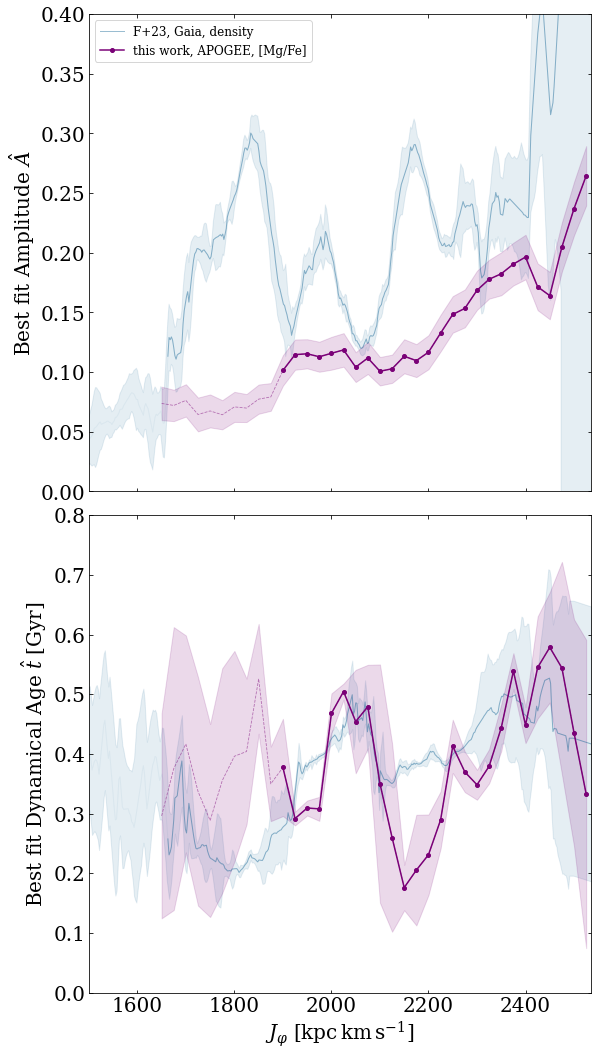}
    \caption{Perturbation amplitude $A$ (eq. \ref{eq:perturbation_amplitude}) and age $t$ as functions of angular momentum. The parameters derived from the [Mg/Fe] Snail structure are shown in solid purple with dots, and those derived from the Gaia phase-space density in \cite{frankel_2023} in light blue. The shaded regions are less than 1-sigma away from the best fits. The best-fit amplitude derived from [Mg/Fe]  
    is somewhat smaller than the amplitude derived from the phase-space density although both curves show similar trends with $J_\varphi$. The best-fit ages span similar ranges as a function of angular momentum. We show low-amplitude results ($A<0.1$) as dashed lines at lower opacity as a reminder that the Iron Snail signal is weak and the best-fit parameters may be unphysical.}
    \label{fig:amp_age_fit}
\end{figure}

\begin{figure*}
    \centering
    \includegraphics[width=\textwidth]{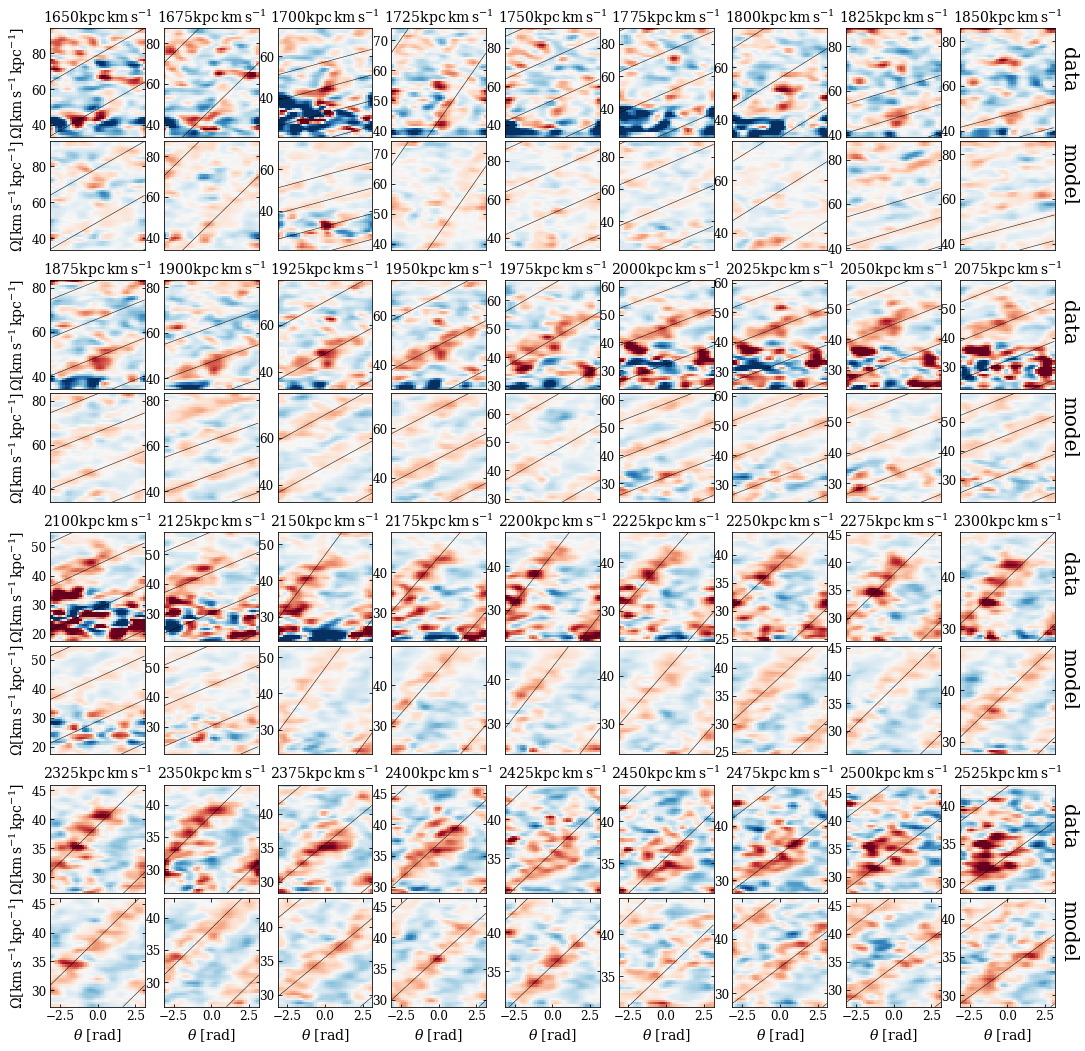}
    \caption{Data vs.\ best-fit model in the vertical angle-frequency plane. In this plane phase-wrapped features should be straight lines; thus the figure is called a Zebra diagram. There are eight rows, where angular momentum increases to the right and to the bottom in bins that are 300 $\mathrm{kpc\,km\,s^{-1}}$ wide and $25~\mathrm{kpc\,km\,s^{-1}}$ apart (i.e., nearby panels are not independent). The tops of the data panels are labeled with the angular-momentum bin centers. The odd rows show data and the even ones show the best-fit model at the same angular momentum, with added noise of 0.03 dex for apple-to-apple comparison. Both data and model panels have been smoothed by the same Gaussian 2D kernel. The black lines show the location of the best-fit model and are also repeated in the data panels to give a comparison anchor. As we can see from the data, the zebra structure is clearly present only for $J_\varphi \gtrsim 1875\,\kpckms.$ In the panels where the data signal is weak, the best-fit model is incoherent, as expected. Where the signal is strong in the data, the best-fit model seems appropriate.}
    \label{fig:enter-label}
\end{figure*}

Here, we discuss the relation and the differences between the measurements of the Gaia Snail (phase-space density) and the Iron Snail. We have argued that the two structures should trace each other as long as there are vertical gradients in the density and the label in the equilibrium disk. The Iron Snail is weaker than the Gaia Snail at low angular momenta, in both the qualitative Figure \ref{fig:Fig 1} and the quantitative measurement in Figure \ref{fig:amp_age_fit}. However, where the perturbation amplitude is large enough ($\gtrsim 0.1$), we find that the two structures mostly agree on the perturbation age, despite the facts that (i) the catalogs used were very different: the former relies on optical space astrometry and derived phase-space densities while the latter relies on a ground-based near-infrared spectroscopic survey and spectroscopic estimates of abundances, (ii) the dynamical analysis is completely different: the former relies on \texttt{Galpy}'s implementation of the Milky Way potential, whereas the latter relies on empirical measurements of orbit shapes with OTI. Nevertheless, the perturbation ages are similar, and the sudden drop in age at $2100\,\kpckms$ is present in both.

However, there are significant differences. The drop in age at $2100\,\kpckms$ is much stronger in the [Mg/Fe]-derived signal. There is a strong oscillatory pattern with angular momentum in the phase-space density-derived amplitude that seems weak or non-existent in the [Mg/Fe]-derived amplitude. 
These differences could mean that our Snail model is too simple. In particular, any realistic impulse should be three-dimensional, and we have only modeled the vertical dimension. We know that phase-space density has strong vertical \emph{and} radial gradients. However, [Mg/Fe] has strong vertical gradients but only weak radial gradients. This could mean that by using the density, we are measuring the net effect of a kick in both dimensions, whereas by using a label that only has vertical gradients, we are measuring the projection of the perturbation in the vertical direction. To test whether this speculation holds is beyond the scope of this work, but using several abundances (e.g., [Fe/H], which has a strong radial gradient and a weak vertical gradient) together could help to disentangle the multidimensional aspect of the dynamical perturbation.

These differences could also reflect the different extent of the disk probed by these two surveys. The  phase-space density based on Gaia data is very local, confined to a radius of 0.5 kpc around the Sun. In contrast, the one on the Iron Snail, based on APOGEE, probes distances of several kpc from us. If the amplitude of the perturbation varies spatially \citep[e.g.,][]{alinder_2023, widmark_2021}, then it is not surprising that the amplitude derived from these two different datasets is different, with the more extended dataset looking smoother.

The orbital parameters for the stars used in determining the Gaia phase-space structure and those  determining the APOGEE abundance structure are different: the former used the Milky Way potential ({\tt MWPotential2014}) as implemented in the \texttt{Galpy} Python package \citep{bovy_2015_galpy}, whereas the latter relies on empirically derived orbital parameters based on the angle-averaged abundances. By construction, the latter procedure yields the strongest residuals in the Snail. But since it empirically derives orbit shapes, and does not enforce the existence of a suitable underlying gravitational potential, it could lead to errors in the vertical frequencies. This issue is discussed extensively in \citet{price-whelan_2024}. Since it is the dependence of vertical frequency on action that sets the dynamical age of a phase mixing perturbation, a disagreement in vertical frequencies between the two methods can yield a disagreement in the dynamical age of the perturbation.

Finally, we note that the perturbation amplitudes inferred from the density and from abundances here are not exactly comparable: the first measurement is at high angular-momentum resolution, with small, independent, angular momentum bins. Due to sparser data in the APOGEE catalog, we here took $300\,\kpckms$-wide bins spaced by $25\,\kpckms$: they are not independent at all, the same star contributes information to several bins, and so we only see a smoothed version of the true signal.

\begin{figure*}
    \centering
    \includegraphics[width=\textwidth]{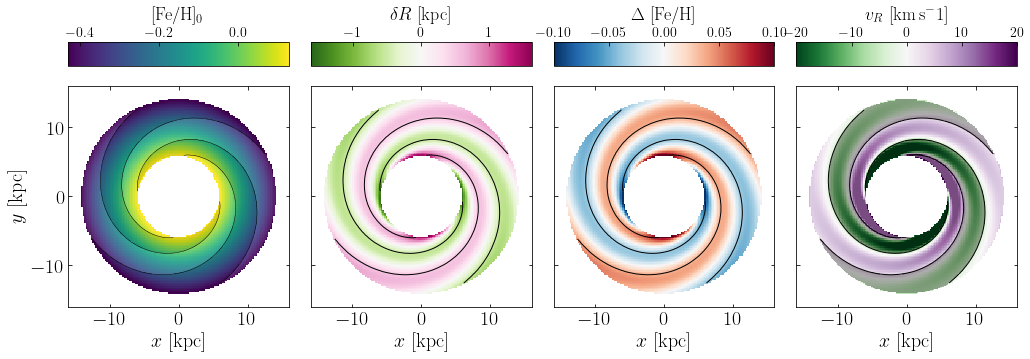}
    \caption{Modeled spiral perturbation in a galactic disk. In all panels, the black contour lines trace the spiral overdensities (in the first and last panel, overdense regions are shaded in a slightly darker color). From left to right: un-perturbed mean metallicity map, perturbation in Galactocentric radius, resulting perturbation in metallicity (stars that moved out are more metal rich, affecting the mean metallicity at a given radius), and finally mean radial velocity map.}
    \label{fig:spiral_patterns}
\end{figure*}

\section{Discussion and Conclusion}

\subsection{Future Extensions to the Snail}

Different labels can (and do) have different profiles in the Galaxy. For example, in the stellar disk, there is a strong radial metallicity gradient, and a weak vertical metallicity gradient. On the other hand, there is a weak radial [Mg/Fe] gradient and a strong vertical [Mg/Fe] gradient, i.e., [Fe/H] and [Mg/Fe] trace almost orthogonal projections of dynamical perturbations (radial vs. vertical). This illustrates that different labels encode different information on the dynamical perturbation of the disk. Exploring the strength of dynamical perturbations in the disk using abundances may open the door to disentangling the various components of dynamical perturbations.

\subsection{Other Applications: spiral density waves in the Galactic Disk}
We have argued that any dynamical perturbation that creates a phase-space density structure should also create a label structure provided that the labels have gradients in phase space. This insight can also be applied to the in-plane motion in the Galactic disk perturbed by spiral overdensities. 

In Gaia and other surveys, the stars coinciding with spiral arms are found to be more metal-rich than stars in inter-arm regions \citep{poggio_2022, hawkins_2023, hackshaw_2024}. A straightforward physical interpretation for this abundance structure could be that the azimuthal variations of mean metallicity are driven by the youngest stars, which are not yet dynamically phase mixed. An alternative interpretation is that the gas metallicity has azimuthal variations, which might occur if chemical enrichment traces star formation, which is more efficient in the spiral arms. Such azimuthal metallicity variations are indeed observed at the $\approx 0.1$ dex level in the gas of the Milky Way disk \citep{balser_2015, wenger_2019}. 
However, \cite{hackshaw_2024} find azimuthal abundance variations in older stars, which this argument \emph{alone} would fail to explain.

A third explanation is that spiral structures create dynamical perturbations that, combined with the radial metallicity gradient in the Galactic disk, produce spiral structure in the mean metallicity. \cite{debattista_2024} have shown in a simulation that spiral abundance structures move with spiral arms at the same pattern speed, and are (in this specific simulation) unlikely to be caused by newly formed metal-rich stars in the spiral arms. \cite{khoperskov_2018A} suggest that the abundance structure results not from a radial metallicity gradient, but from the combination of a vertical metallicity gradient with the fact that dynamically colder (so more metal-rich) populations have a greater response than hotter (more metal-poor) ones to spiral perturbations.

Here, we briefly investigate a simple model for a dynamical perturbation of the Milky Way disk inspired by \cite{eilers_2020}\footnote{With an extra term arising from the logarithmic form of the spirals, which was missing in the original reference.}. We use a spiral model close to their best fit, which produces mean radial velocity patterns with an amplitude of $\sim 10 \mathrm{km\,s^{-1}}$. The model consists mostly of nearly circular orbits that are weakly perturbed by an $m=2$ spiral mode, and is illustrated in Figure \ref{fig:spiral_patterns}. As the third panel of Figure \ref{fig:spiral_patterns} shows, the same dynamical perturbation produces mean metallicity variations at the 0.1 dex level, i.e., the same values as measured in \cite{hawkins_2023, poggio_2022, hackshaw_2024}. Our model does not include vertical kinematics and so it cannot account for the fact that colder (and metal-richer) orbits respond  more strongly to perturbations as suggested in \cite{khoperskov_2018A}. However, this latter scenario should in principle produce a significant signal in [Mg/Fe]\footnote{Or any alpha element, the point here is that those abundances correlate better with age, which correlates better with disk height than with Galactocentric radius.} rather than just in [Fe/H] since the vertical [Mg/Fe] gradient is stronger and clearer than the [Fe/H] one. \citet[][Fig.\ 7]{hackshaw_2024} show that this is actually the case. However, they also find that older stars (2--6 Gyr) which typically have higher velocity dispersions, follow the spiral structure, too (and so probably respond to the spiral perturbation despite being older/hotter). Once again, disentangling the mechanisms governing the response of the disk to spiral perturbations would benefit from quantitatively linking at least two distinct abundance structures to dynamical perturbations.

\subsection{Summary}
When a stellar system undergoes a dynamical perturbation, the subsequent phase mixing of stars will often appear as a spiral structure in phase-space density. The spirality results directly from the presence of frequency gradients in most systems. If there are any gradients between the invariant labels of stars (e.g., abundances) and stellar orbits, then a similar spiral structure should also appear in the distribution of the mean label in phase space. We used the Gaia Snail as an example, and showed that:
\begin{itemize}
    \item the Snail also appears in mean [M/H], [Fe/H], [Mg/Fe], [C/N], age, asteroseismic mass and other labels;
    
    \item the properties of the Snail, in particular age and amplitude, are roughly similar when measured using phase-space density and the abundance ratio [Mg/Fe]. 
    
    \item mean labels, or label moments, have the advantage over phase-space density that they are less sensitive to spatial selection effects. We have shown this by exhibiting the Snail in the results from the APOGEE survey, which has complex spatial selection effects. 

    \item more generally, the same line of reasoning can apply to other types of dynamical perturbations as long as there are gradients between orbits and labels. We showed an example with the metallicity signal that traces spiral patterns in the Milky Way disk.
    
\end{itemize}

\section*{Acknowledgements}
It is a pleasure to thank Teresa Antoja, Jo Bovy, Rimpei Chiba, Jason Hunt, Kathryn Johnston, Juna Kollmeier, Paul McMillan, Josh Speagle and  David Weinberg for interesting discussions.

This project was developed in part at the Gaia Hike, a workshop hosted by the University of British Columbia and the Canadian Institute for Theoretical Astrophysics in 2022 June (\url{https://gaiahike.cita.utoronto.ca/}). 

Funding for the Sloan Digital Sky 
Survey IV has been provided by the 
Alfred P. Sloan Foundation, the U.S. 
Department of Energy Office of 
Science, and the Participating 
Institutions. The SDSS 
website is \url{www.sdss4.org}.

SDSS-IV acknowledges support and 
resources from the Center for High 
Performance Computing at the 
University of Utah.

SDSS-IV is managed by the 
Astrophysical Research Consortium 
for the Participating Institutions 
of the SDSS Collaboration including 
the Brazilian Participation Group, 
the Carnegie Institution for Science, 
Carnegie Mellon University, Center for 
Astrophysics | Harvard \& 
Smithsonian, the Chilean Participation 
Group, the French Participation Group, 
Instituto de Astrof\'isica de 
Canarias, The Johns Hopkins 
University, Kavli Institute for the 
Physics and Mathematics of the 
Universe (IPMU) / University of 
Tokyo, the Korean Participation Group, 
Lawrence Berkeley National Laboratory, 
Leibniz Institut f\"ur Astrophysik 
Potsdam (AIP),  Max-Planck-Institut 
f\"ur Astronomie (MPIA Heidelberg), 
Max-Planck-Institut f\"ur 
Astrophysik (MPA Garching), 
Max-Planck-Institut f\"ur 
Extraterrestrische Physik (MPE Garching), 
National Astronomical Observatories of 
China, New Mexico State University, 
New York University, University of 
Notre Dame, Observat\'ario 
Nacional / MCTI, The Ohio State 
University, Pennsylvania State 
University, Shanghai 
Astronomical Observatory, United 
Kingdom Participation Group, 
Universidad Nacional Aut\'onoma 
de M\'exico, University of Arizona, 
University of Colorado Boulder, 
University of Oxford, University of 
Portsmouth, University of Utah, 
University of Virginia, University 
of Washington, University of 
Wisconsin, Vanderbilt University, 
and Yale University.

This work has made use of data from the European Space Agency (ESA) mission Gaia (\url{https://www.cosmos.esa.int/gaia}), processed by the Gaia Data Processing and Analysis Consortium (DPAC, \url{https://www.cosmos.esa.int/web/gaia/dpac/consortium}). Funding for the DPAC has been provided by national institutions, in particular the institutions participating in the Gaia Multilateral Agreement.

Software: Numpy \cite{numpy}, Astropy \citep{astropy}, JAX \citep{jax2018github},
JAXOpt \citep{jaxopt_implicit_diff}, Matplotlib \citep{matplotlib}, Scipy \citep{scipy}, Galpy \citep{bovy_2015_galpy}, torusimaging \citep{price-whelan_2024}.

NF acknowledges the support of the Natural Sciences and Engineering Research Council of Canada (NSERC), funding reference numbers 568580 and RGPIN-2020-03885, and partial support from an Arts \& Sciences Postdoctoral Fellowship at the University of Toronto. 

JS acknowledges the support of the Natural Sciences and Engineering Research Council of Canada (NSERC), funding reference number 587652.

This Iron Snail project was carried on a 1,200 km cycling network connecting CITA to CCA, including the Empire State Trail.

\bibliography{lit}

\end{document}